\documentclass[12pt]{iopart}

\usepackage{graphicx}
\usepackage{bm}
\usepackage{textcomp}
\usepackage{color}
\usepackage{amsfonts}
 
\newcommand{\red}{\textrmcolor{red}}

\newcommand{\rem}[1]{}
\begin{document}

\title{Edge states at the boundary between topologically equivalent lattices}
\author{Helena Massana-Cid$^{a,d}$, Adrian Ernst$^{a}$,  Daniel de las Heras$^{a}$, 
	Adam Jarosz$^b$, Maciej Urbaniak$^b$, Feliks Stobiecki$^b$, 
	Andreea Tomita$^c$, Rico Huhnstock$^c$, Iris Koch$^c$, Arno Ehresmann$^c$, Dennis Holzinger$^c$, and Thomas M. Fischer$^{a*}$}
\address{$^{a}$ Institute of Physics and Mathematics, Universit\"at Bayreuth, D-95440 Bayreuth (Germany).
	$^b$ Institute of Molecular Physics, Polish Academy of Sciences, 
	ul. M. Smoluchowskiego 17, 60-179 Pozna\'n (Poland).
	$^c$ Institute of Physics and Centre for Interdisciplinary Nanostructure Science and Technology (CINSaT), Universit\"at Kassel, Heinrich-Plett-Strasse 40, D-34132 Kassel (Germany)
	$^d$ 
	Departament de F\'isica de la Mat\`eria Condensada, Universitat de Barcelona, Avinguda Diagonal 647, 08028 Barcelona, (Spain)}

$^*$email:thomas.fischer@uni-bayreuth.de

	\vspace{10pt}
\begin{indented}
	\item[]\date{\today}
\end{indented}

\begin{abstract}{
Edge currents of paramagnetic colloidal particles propagate at the edge between two topologically
equivalent magnetic lattices of different lattice constant when the system is driven with periodic
modulation loops of an external magnetic field. The number of topologically protected particle edge transport
modes is not determined by a bulk-boundary correspondence. Instead, we find a rich variety of edge transport modes
that depend on the symmetry of both the edge and  the modulation loop. The edge transport can be ratchet-like or adiabatic, and time
or non-time reversal symmetric.
}\end{abstract}

\maketitle
\section{Introduction}

The bulk-boundary correspondence states that two "wave" systems coupled at an edge will support edge-states
at the boundary that live in the frequency gap of both bulk systems provided that the Chern numbers
of the upper bands of both coupled systems are different~\cite{Hasan}. Rudner, Lindner, Berg and Levin~\cite{Rudner}
have generalized the bulk-boundary correspondence to periodically driven two-dimensional systems. There, 
the number of edge modes is characterized by the difference of the winding number of the time dependent
evolution operators of the two bulk systems coupled at the edge. Hence, the bulk properties of the coupled
system are sufficient to predict the existence or absence of edge states. No knowledge of the edge between the two bulk systems
is required.

In previous works we have shown how the motion of colloidal particles above periodic magnetic lattices
can be of topological nature~\cite{tp1,tp2,tp3,colloidalTI}, and how the same topological concepts apply
to both particle and "wave" systems~\cite{Rechtsman,Perczel,Kane,Mao,Paulose,Nash,Huber,Murugan}.
Working with colloidal systems, as opposed to "wave" systems has however the advantage of a direct visualization of the motion 
and can therefore be more intuitive. 

Here, we present an example of propagating edge colloidal currents between two
coupled bulk lattices that differ only in their lattice constant and therefore are
topologically equivalent.

Since we couple a primitive unit cell of one
lattice to a larger unit cell of the other lattice, the translation invariance of the larger lattice along the edge
is preserved in the coupled system. There exists no continuous deformation of one lattice to the other lattice that
preserves this translation invariance. Therefore, the bulk-edge correspondence does not apply here but we still find
propagating edge states despite the topological equivalence between both lattices. The edge states are a robust
feature since they are topological in nature. However, they are topological not because of a contrast in topology
between two bulk materials. The edge states are edge penetrating spirals that share similarities with skipped
orbits \cite{Beenakker,Davies,Shi,Montambaux,Zhirov,Mancini,colloidalTI} that are edge states between topologically
distinct lattices. The response of the edge current to the driving force differs depending on whether the winding numbers of the 
modulation loop around special points are odd or even. We demonstrate that a variety of edge
transport modes is possible between topologically equivalent lattices.

\section{Edge transport between two square lattices}
\begin{figure}
		\includegraphics[width=0.95\columnwidth]{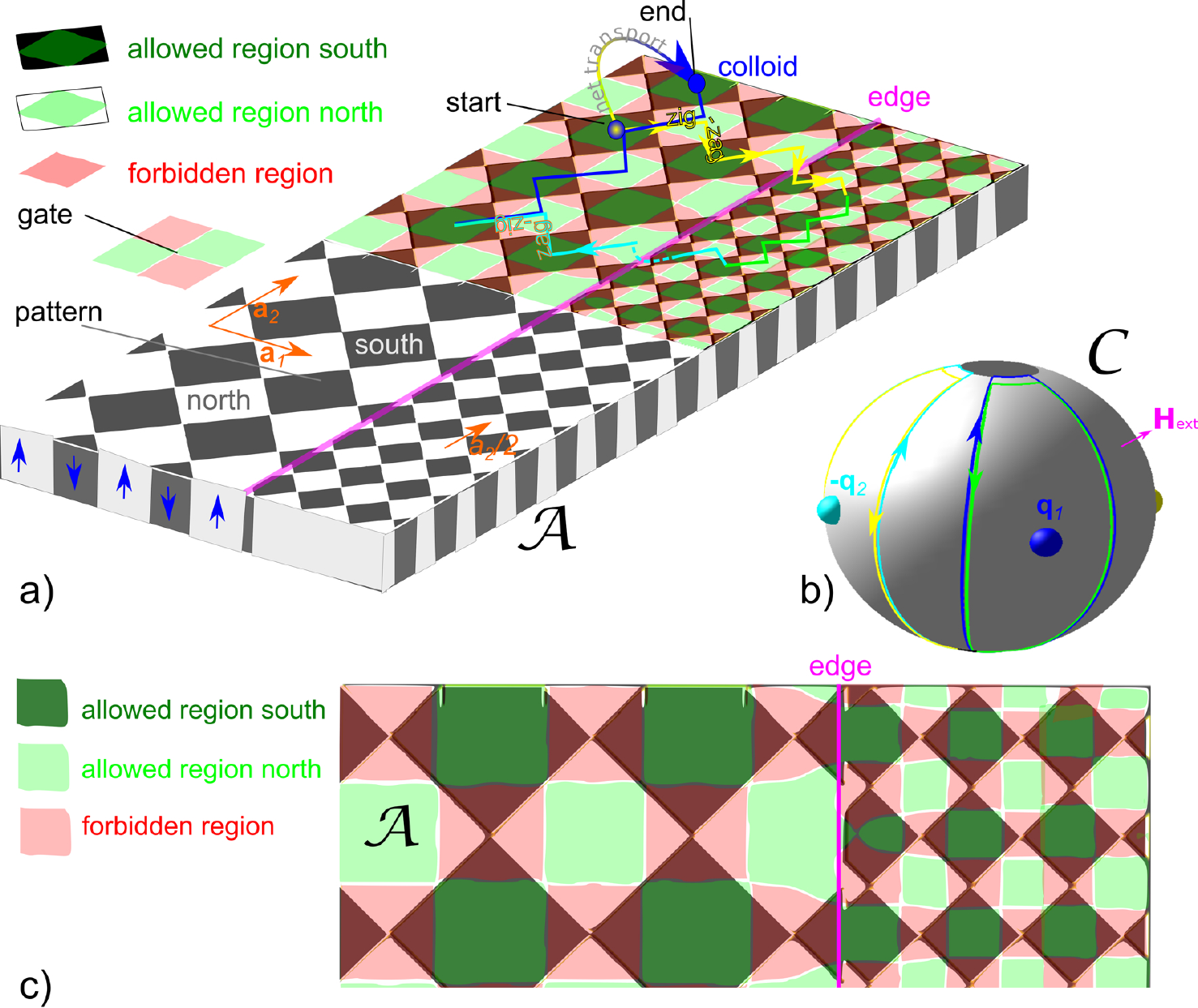}
	\caption{a) A paramagnetic colloidal particle on top of a magnetic artificial domain pattern subject to a time dependent superposed external magnetic field $\bf H_{\textrm{ext}}(t)\in{\cal C}$ of constant strength but varying orientation.
   The magnetic pattern below the colloid  consists of two square lattices of lattice constants $a$ and $a/2$ merging at an edge. The regions above the pattern that can be converted into a minimum of the colloidal potential are the allowed regions, the other regions are forbidden and they can only be converted into saddle points of the colloidal potential but not into extrema. A loop in control space can cause a net transport over a period at the edge even when it does not cause net motion in the bulk of either lattice. The chiral cyclotron control loop shown in b) adiabatically transports the colloid on a zig-zag path (yellow and green)  and zag-zig path (cyan and blue) through allowed regions until it hits a forbidden region and performs a ratchet jump (dashed cyan line) and then continuous through allowed regions. At the end of the loop the colloid is displaced by a large unit vector along the edge. c) an example of the allowed and forbidden regions for a particular symmetric edge pattern.}
	\label{fig1}
\end{figure}

We illustrate the richness of the edge transport modes between topologically equivalent lattices using the simplest lattice
that supports topological transport in bulk: a square magnetic lattice~\cite{tp2,tp3}. We study experimentally and 
with computer simulations the colloidal motion above the edge of two joint magnetic square patterns with different
size of the unit cell. The primitive lattice vectors of the large pattern double in magnitude the primitive lattice vectors of the small pattern.
In the experiments, paramagnetic colloidal particles move above a thin 
Co/Au layered system with perpendicular magnetic anisotropy 
lithographically patterned via ion bombardment~\cite{CBF1998,KET2010,tp3}. The pattern consists
of a patch of alternating square domains with a mesoscopic pattern lattice constant $a\approx 14\,\mu\textrm{m}$
adjacent to a half sized square pattern, see a sketch in Fig. \ref{fig1}a. The whole pattern is magnetized in the
$\pm z$-direction normal to the film. The magnetic pattern is spin coated with a $1.6\, \mu m$ polymer film that serves as a spacer. The paramagnetic colloidal particles (diameter
$2.8\,\mu\textrm{m}$) are immersed in water and their centers move
at a fixed elevation $z\approx 3\mu m$ above the pattern~\cite{tp3}.  Hence, the colloidal particles move in a two-dimensional
plane that we refer to as action space, $\cal A$. A uniform time-dependent
external magnetic field $\bf{H}_{\textrm{ext}}(t)$ of constant magnitude ($H_{\textrm{ext}}=4\,\textrm{kAm}^{-1}$)
is superimposed to the non-uniform and time-independent magnetic field generated by the pattern $\bf{H}_{\textrm{p}}$.
We also study the system using Brownian dynamics simulations. Details of the simulation are provided in the appendix.

The external field $\bf{H}_{\textrm{ext}}(t)$ is varied on the surface of a sphere that we call the
control space $\cal C$ (see Fig. \ref{fig1}b). We perform periodic modulation loops $\cal L_C$ of the external field in control space
to drive the system. Special modulation loops in control space induce colloidal motion in action space in the bulk of both square patterns
and at the edge between them.

\begin{figure}
	\includegraphics[width=0.9\columnwidth]{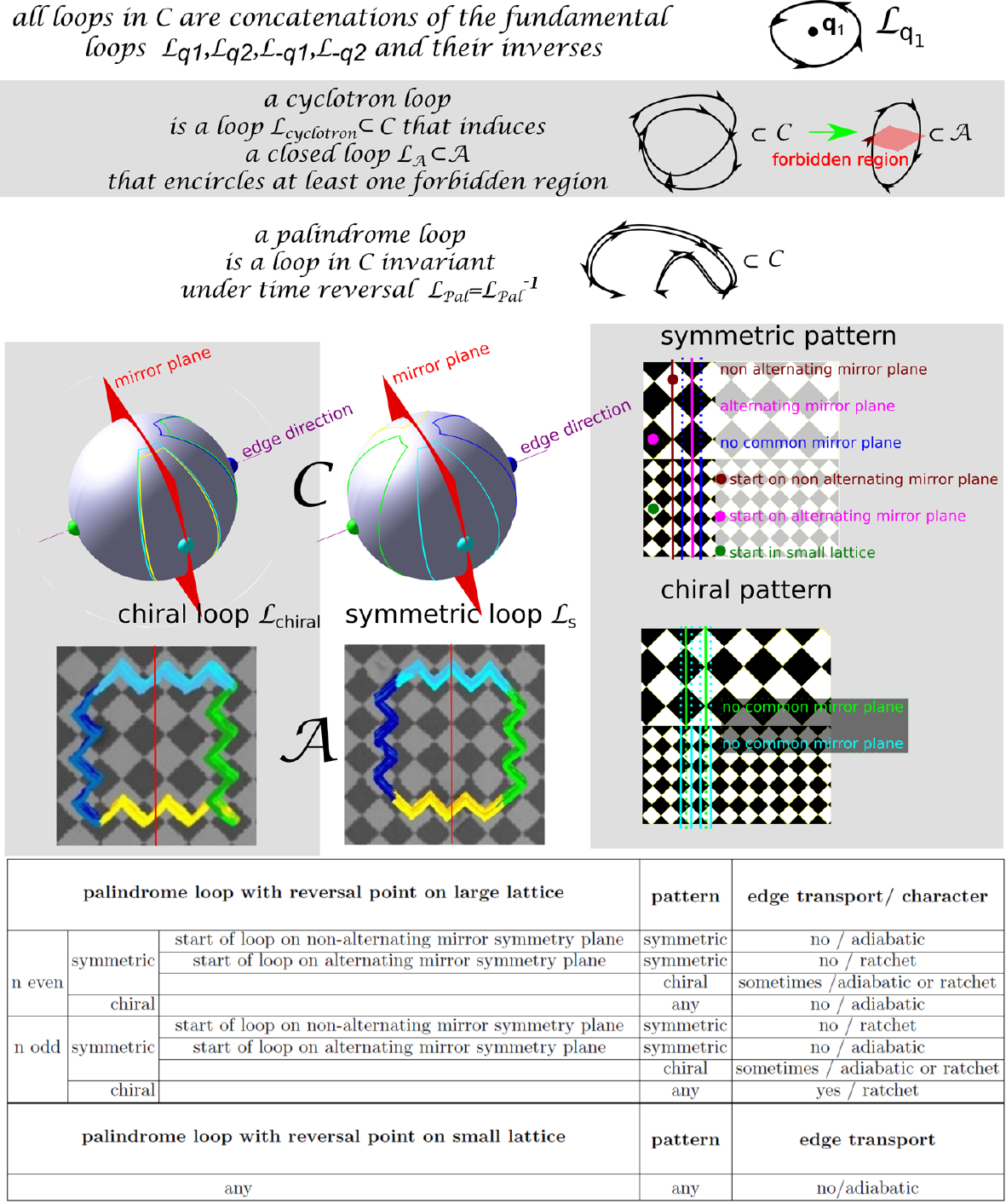}
	\caption{Explanation of the various types of loops in control space, of the types of patterns and a table of resulting edge transport in action space ${\cal A}$.}\label{definitions}
\end{figure}

In Refs.~\cite{tp1,tp2,tp3} we demonstrated how the bulk transport of colloidal particles
above magnetic lattices with different symmetries is topologically protected. We summarize here
the main aspects of the transport in square lattices and refer the reader to Refs.~\cite{tp2,tp3} for a complete description.
For each lattice symmetry there exist special modulation loops of $\bf{H}_{\textrm{ext}}$ in $\cal C$ that induce transport of
colloids in $\cal A$. These loops share a common feature, they wind around special objects in $\cal C$~\cite{tp1,tp2,tp3}.
In the simplest case, a square lattice, the control space is characterized by just four "fence" points on the equator
lying along the directions of the smallest reciprocal lattice vectors~\cite{tp2}, see Fig~\ref{fig1}b.  A modulation
loop encircling one of these points ($\bf q_1$,$\bf q_2$,-$\bf q_1$, -$\bf q_2$) in $\cal C$ in the
mathematical positive sense  transports the colloids in $\cal A$ one unit cell along 
one of the four possible directions of the square lattice $-\bf {a}_2,\bf {a}_1,\bf {a}_2, -\bf {a}_1$. We call a modulation loop encircling one of the fence points a fundamental loop.
Each of these fundamental loops induces adiabatic transport in the sense that the colloidal particle follows a minimum
of the colloidal potential at any time. Hence, the position of the particle in ${\cal A}$  parametrically depends on the
position of the loop in ${\cal C}$. All modulation loops discussed in what follows can be viewed as concatenations of fundamental loops. A summary of such loops is given in figure \ref{definitions}.

Action space can always be split into allowed and forbidden regions~\cite{tp1,tp2,tp3}. Any point inside the allowed (forbidden) regions 
can be rendered into a minimum (saddle point) of the potential with a suitable choice of the external field.
The allowed and forbidden regions in square patterns form another checker square pattern turned by $45$ degrees with respect to the
magnetization pattern~\cite{tp2}. The allowed squares are centered above the domains of the pattern and the forbidden regions are centered
above the domain wall crossings, see Fig.~\ref{fig1}a. The allowed regions can be further split into south and north (Fig. 1). The colloidal particle is located in the allowed region north (south) when the external field points into the northern (southern) hemisphere.

Two adjacent allowed regions in ${\cal A}$ meet at one point that we refer to as the gate (Fig~\ref{fig1}a). In control space the gates are
segments of the equator joining two consecutive fence points. When the external field passes a gate in control space then
the particle moves through the corresponding gate in action space. A modulation loop winding around a fence point passes two gates and the resulting
transport in action space is a zig-zag move through both gates with the zig and zag at 45 degrees to the unit vectors of the pattern (Fig. 1a). 
An example of such loops in ${\cal A}$ and ${\cal C}$ is schematically shown in Fig.~\ref{fig1} panels (a) and (b), respectively.
Colloidal particles can be adiabatically moved into the positive $\bf {a}_1$-direction 
either in a zig-zag or in a zag-zig way avoiding a neighboring forbidden region on different sides (right respectively left)
by a counterclockwise winding of the control loop ${\cal L}_{\bf q_2}$ around the $\bf q_2$ fence point of control space, and
into the same positive $\bf {a}_1$-direction by a 
a clockwise winding  ${\cal L}_{-\bf q_2}^{-1}$ around the $-\bf q_2$
fence. Both fundamental moves are topologically distinct mirror images of each other
${\cal L}_{\bf q_2}=\sigma_2({\cal L}_{-\bf q_2}^{-1})$ since they wind around different fence points. Also the resulting trajectories in $\cal A$ are topologically distinct since they pass the
forbidden region in action space on different sides they cannot be continuously deformed into each other. Here $\sigma_i$ is the mirror
reflection $\bf a_i\to -\bf a_i$  (and hence $\bf q_i\to -\bf q_i$).
The mirror image of the loop ${\cal L}_{\bf q_1}=\sigma_2({\cal L}_{\bf q_1}^{-1})$ is equal to its time reversed loop around the same fence point $\bf q_1$. The bulk concept of allowed and forbidden regions as well as the concept of gates can
be generalized to the region of the edge. Here the allowed and forbidden regions become distorted and the number of gates
might be different from the bulk depending on the fine details of the edge.  An example of the allowed and forbidden regions
close to the edge is shown in Fig. \ref{fig1}c.

\section{Edge transport with cyclotron control loops}    

In the bulk of a metal electrons perform cyclotron orbits if they are exposed to a constant magnetic field. These cyclotron orbits are loops enclosing multiples of flux quanta. In analogy to such electrons we call a modulation loop in $\cal C$ a cyclotron modulation loop if the colloids in the bulk of $\cal A$ perform a loop encircling an integer number of forbidden regions (see Fig.\ref{definitions}).  
Combining the fundamental bulk transport modes we may construct cyclotron orbits of square side length $na$,
$n=1,2,3...$ of a paramagnetic particle above the bulk of a square lattice. The corresponding cyclotron control
loop in $\cal C$ consists of four connected parts. Each of the four parts of the loop
winds $n$ times around a fence point of $\cal C$, and therefore transports a particle $n$ unit cells
along one of the lattice vectors in ${\cal A}$. There are various types of cyclotron loops. A symmetric cyclotron loop (Fig. \ref{definitions})
${\cal L}_{Cycl,s}={\cal L}_{-\bf q_2}^{-n}{\cal L}_{-\bf q_1}^{-n}{\cal L}_{\bf q_2}^{-n}{\cal L}_{\bf q_1}^{-n}$ 
 is a loop where the mirror image 
of the endlessly repeated loop $\lim_{m\to\infty}{\cal L}_{Cycl,s}^m$ coincides with the time reversed endless loop, i.e.,
$\sigma_2({\cal L}_{Cycl,s}^\infty)={\cal L}_{Cycl,s}^{-\infty}$. We also work with chiral cyclotron loops 
${\cal L}_{Cycl,chiral}={\cal L}_{\bf q_2}^n{\cal L}_{\bf q_1}^n{\cal L}_{\bf q_2}^{-n}{\cal L}_{\bf q_1}^{-n}$
(see Fig. \ref{definitions})  for which the mirror image of the endless repeated loop differs from the time reversed endless loop.

Besides the symmetry of the modulation loop we also must consider the symmetry of the pattern. 
There are patterns with mirror symmetry along a plane perpendicular to the edge.
We refer to these patterns as symmetric patterns. We distinguish non alternating mirror
planes which run across equivalent square domains of the same magnetization in the large and small lattice (see Fig.~\ref{definitions})
from alternating mirror planes that run across squares of opposite magnetization on the large and small lattice (Fig.~\ref{definitions}).
Both types of mirror planes cover equivalent squares on the smaller lattice of only one magnetization such that there are no common mirror planes on the
squares of the oppositely magnetized squares in the smaller lattice. 
When the symmetry planes of the small and large lattices do not match we call the pattern a chiral pattern. One chiral pattern is depicted in Fig. \ref{definitions}. 

The symmetries of the modulation loops and of the pattern are crucial for understanding the transport of colloids at the edge of both lattices. Combinations of different symmetries cause a rich variety of transport phenomena which we have summarized in the table at the bottom of Fig. \ref{definitions}.

\begin{figure}
	\includegraphics[width=0.95\columnwidth]{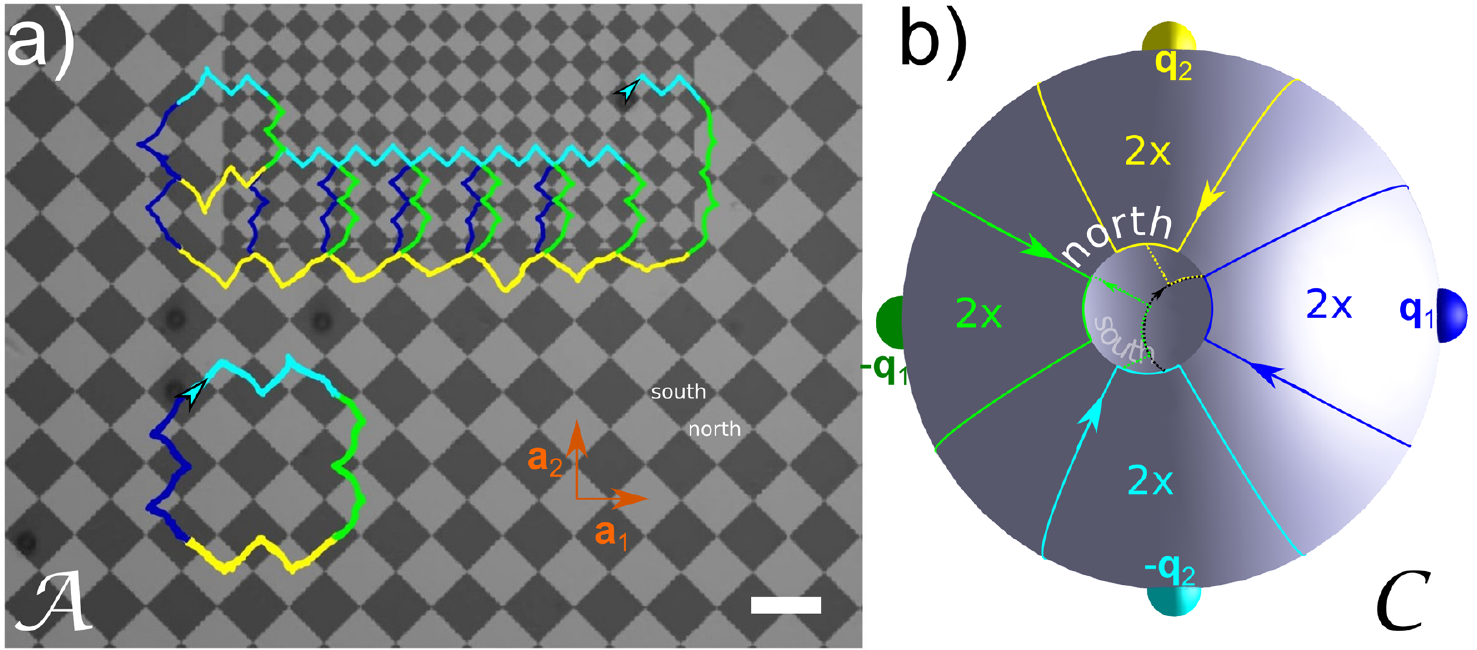}
	\caption{ {\bf a)} Experimentally determined bulk cyclotron orbit and penetrating edge spiral path of
		two colloidal particles on the square pattern in action space $\cal A$ (the pattern is overlayed to the microscope image since it can be visualized only with very low contrast) for {\bf b)} the symmetric cyclotron orbit ${\cal L}_{Cycl,s}(n=2)$ in $\cal C$. The fundamental loops each of which is repeated twice are shown in colors corresponding to the segments in a). Fundamental loops start and end in the south where they are concatenated to the next fundamental loop. Fence points (encircled spheres) are shown in equivalent colors. The edge current surrounds two corners of the pattern. Scale bar is 14 $\mu m$. A video clip of the motion of the paramagnetic colloidal particles is provided in $adfigure3.avi$.}\label{fig2}
\end{figure}

Let us start with a symmetric cyclotron loop.
In Fig. \ref{fig2}b we show a ${\cal L}_{Cycl,s}(n=2)$ in $\cal C$.
The corresponding colloidal trajectory in $\cal A$ is depicted in Fig.~\ref{fig2}a.
Colloidal particles above the bulk of the large square pattern perform closed 
cyclotron orbits of the imposed side length ($n=2$ in this example). Control space is the same
for both, the large and the rescaled small pattern. Therefore, the cyclotron modulation loop also winds around the
fence points in $\cal C$ of the rescaled pattern.
Hence, the colloids above the rescaled square pattern perform a rescaled cyclotron orbit.
The radius of the orbit simply adjusts to the new scale of the pattern. Cyclotron orbits that cut through the edge
have a large radius on the large pattern and a small radius on the small pattern. As a result, the starting and the
ending point of the orbit is shifted by a lattice vector along the edge of the pattern. 
If we start on the large pattern side of the edge, then we end on the large pattern side but displaced
by a large pattern unit vector from the starting point. Hence a particle near the edge driven with a $2$-cyclotron-loop
${\cal L}_{Cycl,s}$ performs an open trajectory along the edge, while the bulk-particles perform
closed cyclotron orbits  (see Fig. \ref{fig2}a and the video clip $adfigure3.avi$). 

\begin{figure}
	\includegraphics[width=0.95\columnwidth]{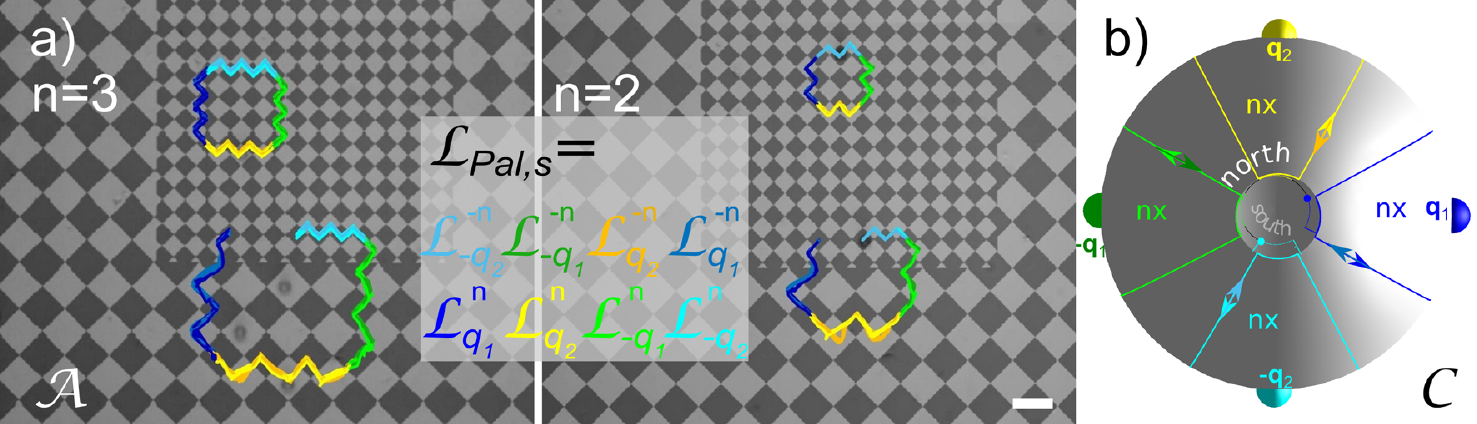}
	\caption{{\bf a)} Experimental trajectories of paramagnets on a pattern subject to a symmetric palindrome cyclotron control loop that reverses after the blue and cyan loop inside the small lattice. Both bulk and edge transport are adiabatic, however the bulk trajectory closes after half of the palindrome loop, while the edge trajectory reverses course at two different points. Scale bar is 14 $\mu m$. {\bf b)} Schematic of the symmetric palindrome cyclotron control loop in control space. Video clips of the motion of the paramagnetic colloidal particles are provided in $adfigure4\_3.avi$ and $adfigure4\_2.avi$. }
	\label{small}
\end{figure}

We have shown in Ref. \cite{tp2} that the bulk transport is entirely adiabatic, i.e. the particle sits during the entire
modulation loop in a potential minimum that is translated in action space. Experimentally, we can show the adiabaticity
or non-adiabdicity by using palindrome cyclotron control loops
${\cal L}_{Pal}={\cal L}_{Cycl}{\cal L}_{Cycl}^{-1}$ where an initial cyclotron control loop is followed by its time reversed loop. A palindrome loop is a loop that is invariant under time reversal (see Fig. \ref{definitions}).
If we infinitely repeat a palindrome loop ${\cal L}_{Pal}^\infty$ we still are able to indentify the reversal points where ${\cal L}_{Cycl}$ and ${\cal L}_{Cycl}^{-1}$ are concatenated. The positions in $\cal A$ of the reversal points are important for the edge transport and different reversal points are depicted in Fig. \ref{definitions}.  
The time reversed loop ${\cal L}_{Cycl}^{-1}$ must undo all the effects of the initial loop ${\cal L}_{Cycl}$ if the transport is adiabatic. Since the bulk transport is
entirely adiabatic~\cite{tp2} the adiabaticity or non-adiabaticity of the edge transport is  decided during the penetration of the edge. 

The smaller lattice has four times the density of minima of the total magnetic potential than the larger lattice,
and each large unit cell at the edge has two smaller neighboring unit cells of the smaller lattice. 
Hence, in a generic situation only one of the minima of the two smaller primitive unit cells can be transferred adiabatically into the
minimum of the large pattern when crossing the edge. The other minimum must be annihilated by a saddle point of one of the smaller
primitive unit cells to create a minima distribution above the large primitive unit cell with half the amount of minima, maxima and saddle
points than the two smaller primitive unit cells. A particle transported via the annihilating small unit cell minimum must
therefore jump along the path of steepest descent toward the minimum of the large primitive unit cell. This is a ratchet jump
and it occurs generically  when exiting the smaller lattice toward the larger lattice. Only for some particular edge patterns there exists also
a jump when crossing the edge from the large to the small lattice.
Entry and exit of the smaller lattice are equivalent when using a chiral cyclotron loop ${\cal L}_{Cycl,chiral}$ that uses a time reversed
fundamental entry loop as exit loop. 
However, if we drive the motion with a palindrome loop it matters whether the point of time reversal lies within the small or within the large lattice.
If the point of return lies within the small lattice then there will be no transport over a period because the entry into the small lattice was
adiabatic and the particle will reverse in the small lattice on the same path and therefore find the adiabatic exit.
The motion is adiabatic upon entry and exit and hence we have no displacement of the particle.
In Fig. \ref{small} we show such adiabatic path for a symmetric palindrome cyclotron control loop of side length $n=2$ and $n=3$.
In both cases the bulk trajectories of the time reversed loop just follow the forward path but in the opposite direction.
In the bulk the two reversal points between  ${\cal L}_{Cycl}$ and  ${\cal L}_{Cycl^{-1}}$ coincide such that already the first cyclotron loop closes the bulk trajectory.
In contrast, at the edge half of the trajectory is an open path with two distinct ends that is reversed in the second half of the modulation loop. On the other hand, if the point of time reversal lies within the large lattice there can be transport in the edge, depending on the edge and loop symmetry. This is discussed in the next section.

 The edge transport cannot generically be adiabatic: 
If the transport were entirely adiabatic then for a cyclotron control loop of side length $n$ that enters and exits the smaller lattice on equivalent paths one would expect the displacement to be $na/2=na-na/2$, i.e $n$-times the difference of the large and small unit vector along the edge. The displacement however, has to be an integer multiple of the large lattice constant $a$. The edge transport can therefore only be adiabatic for $n$ even and must be of the ratchet type for such cyclotron control loops with $n$ odd. 

\section{Edge transport with palindrome cyclotron control loops on symmetric patterns}

\begin{figure}
	\includegraphics[width=0.95\columnwidth]{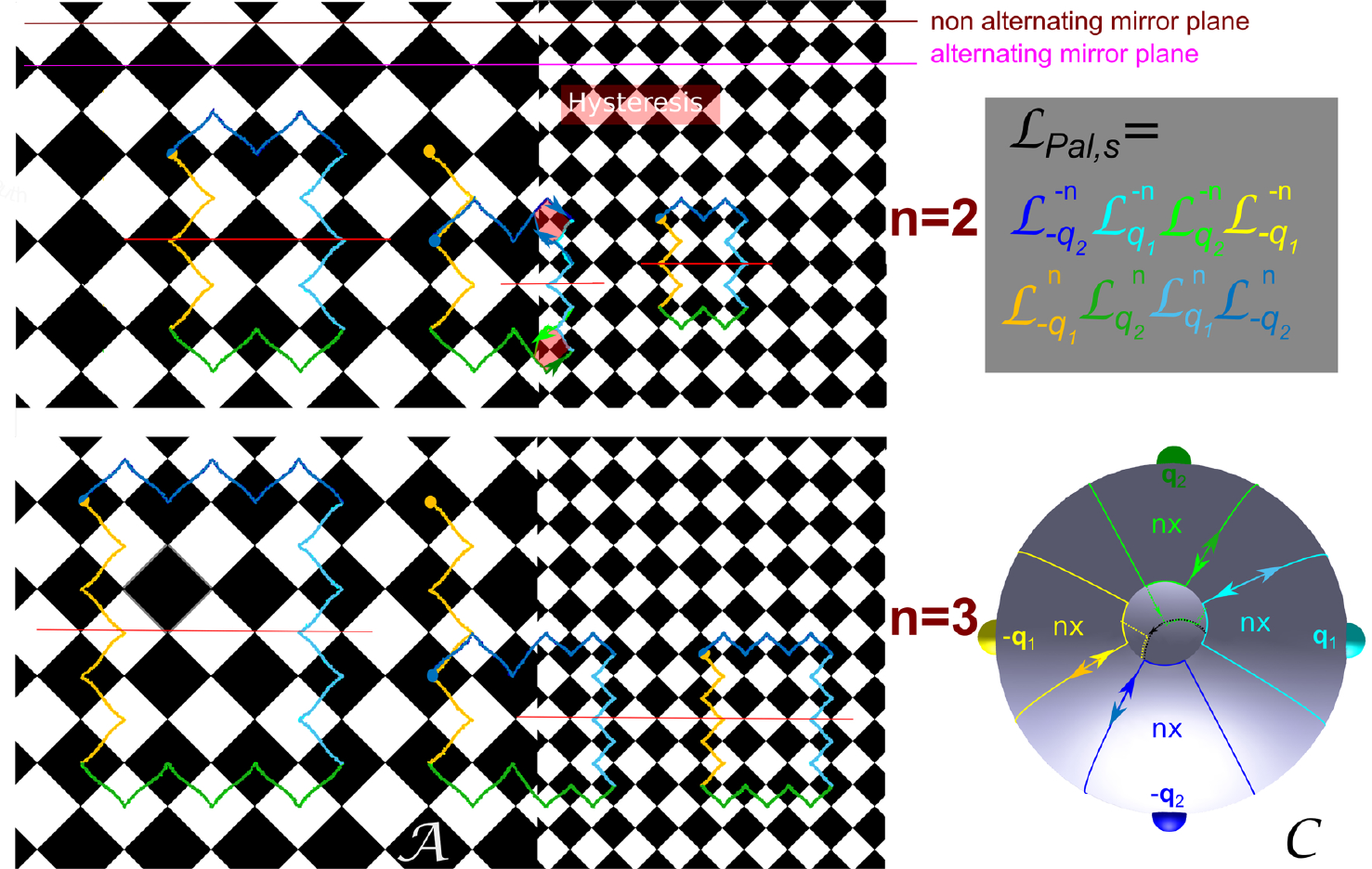}
	\caption{Trajectories of paramagnets obtained by Brownian Dynamics simulations on an symmetric pattern subject to symmetric palindrome cyclotron control loops. Here we show the situation when the fundamental loops building up the control loop start and end in the south of control space (on a black square in action space). Even loops cause ratchet motion with hysteresis and odd loops are fully adiabatic in this case. The loops start and end on squares that have the opposite magnetization of the non-alternating mirror planes. Video clips of the simulations of the paramagnetic colloidal particles are provided in $adfigure5\_3.avi$ and $adfigure5\_2.avi$.}
	\label{fig3}
\end{figure}

Let us first consider patterns with mirror symmetry along a plane perpendicular to the edge (see Fig. 2).
 Due to their symmetry, symmetric loops enter the small lattice on either
non-alternating or on alternating mirror planes depending on whether the fundamental loops used for the symmetric loop start and
end on squares having the magnetization of the mirror planes or not. On a non-alternating mirror plane the zig-zag path can
penetrate into the small lattice without displacement along the edge. In contrast, on an alternating mirror plane the zig-zag
path must leave the alternating mirror plane upon entry into the smaller lattice to a small lattice mirror plane that is not
a mirror plane of the large lattice and will be displaced along the edge by $a/4$. This causes odd and even symmetric loops to behave
the same way as chiral loops when the fundamental loops start and end on squares having the magnetization of the non-alternating mirror planes.
If the fundamental loops however start on the squares with alternating mirror symmetry then symmetric palindrome loops behave opposite to the
chiral palindrome loops and will be adiabatic when $n$ is odd and of the ratchet type for $n$ even Fig. \ref{fig3}. 

\begin{figure}
	\includegraphics[width=0.95\columnwidth]{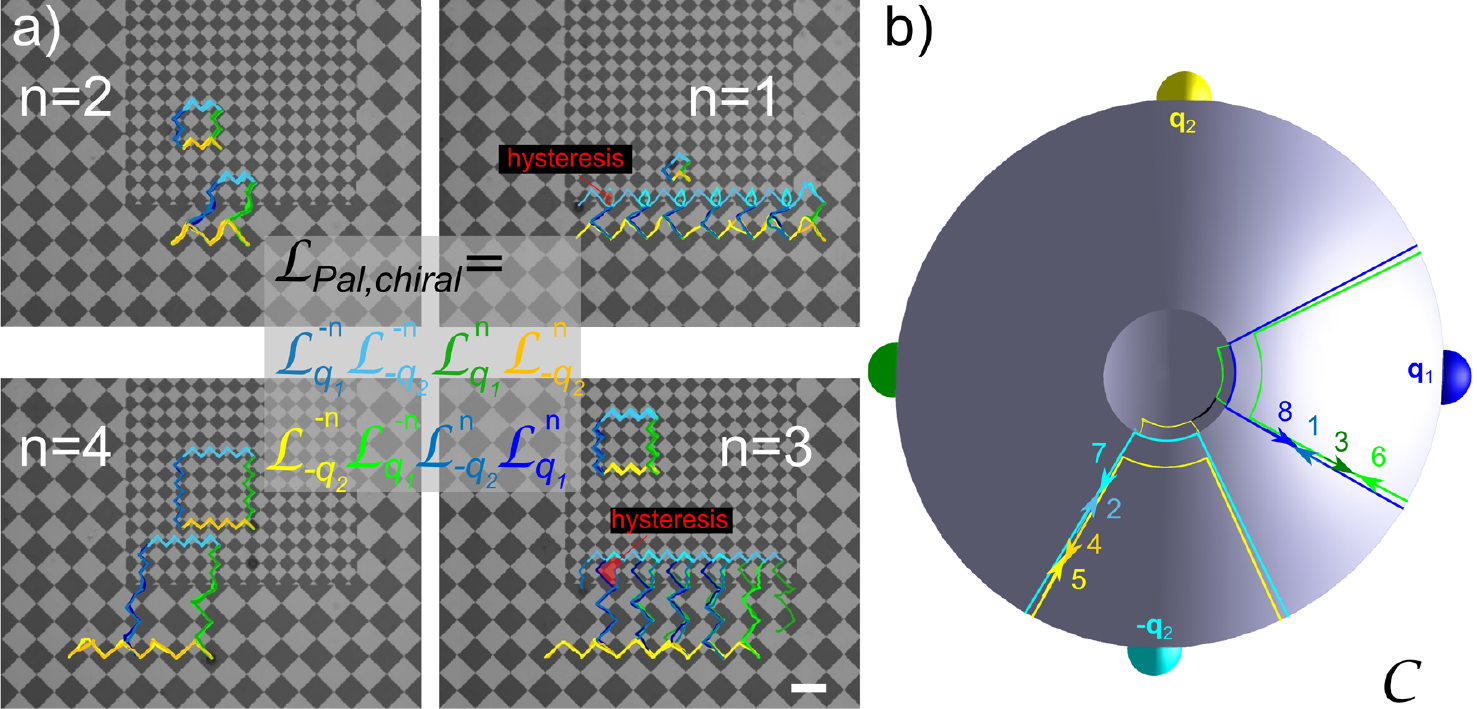}
	\caption{ {\bf a)} Experimentally determined bulk chiral palindrome closed orbit and edge path of two colloidal particles on the square pattern in action space $\cal A$ for {\bf b)} the chiral palindrome orbit ${\cal L}_{Pal,chir}(n)$ in $\cal C$. The fundamental loops are shown in colors corresponding to the segments in a). Fence points (encircled spheres) are shown in equivalent colors. The edge paths are closed for $n$ even and open penetrating spirals for $n$ odd. The adiabatic entry (dark blue) and the ratchet exit (blue of step 8) path differ and enclose a hysteresis (red). The same is true for steps 6 and step 3 of the palindrome loop. Scale bar is 14 $\mu m$. Video clips of the motion of the paramagnetic colloidal particles are provided in $adfigure6\_1.avi$ -  $adfigure6\_4.avi$.}
	\label{fig4}
\end{figure}

Symmetric even loops cause cyclotron orbits in the bulk that are symmetric around a line of symmetry covering squares of the same magnetization as the start of the loop. The edge is symmetric around only one type of smaller squares in the small lattice. Whenever the
small squares of the line of symmetry match the magnetization of the start square, the transport of the symmetric loops
is adiabatic, whenever they cover different types of small squares the transport is of the ratchet type.  This result is
a direct consequence of coupling a large primitive unit cell of the large lattice with two primitive unit cells of the smaller
cell. In Fig. \ref{fig3} and in the video clips $adfigure5\_2.avi$ and $adfigure5\_3.avi$ we show simulated trajectories of "bulk
particles" and particles near a symmetric edge with mirror planes running across the white squares of the small lattice. The
edge  runs in the $\bf a_2$-direction and the colloids are subject to a mirror symmetric palindrome control loop
${\cal L}_{Pal,s}$ of side length $na$ with fundamental loops starting and ending on black squares. Trajectories of bulk
particles are closed in both lattices no matter whether $n$ is odd or even. At the edge the trajectories follow the reversed
path for $n$ odd but have regions where the reverse path is different from the initial path when $n$ is even. Ratchet jumps
at the forward exit from the small lattice are compensated by mirror symmetric ratchet jumps at the reverse exit. We may
however use odd chiral palindrome control loops ${\cal L}_{Pal,chiral}(n=\textrm{odd})$ where both ratchet jumps do not
compensate but cause a net transport along the edge. Experimental trajectories together with the chiral palindrome cyclotron
loop are shown in Fig. \ref{fig4} and in the video clips $adfigure6\_1.avi$ -  $adfigure6\_4.avi$.

\section{Edge transport with palindrome cyclotron control loops on chiral patterns}

Instead of breaking the mirror symmetry $\sigma_2$ by driving with a chiral palindrome loop one can
also produce chiral patterns (see an example of a chiral pattern in Fig. 2). Each of the two  square lattices has mirror reflection planes that
run perpendicular to the edge. If the mirror planes of one lattice
do not coincide with the mirror planes of the other lattice we refer to the pattern as chiral.
The behavior of particles near the edge
is the same as for non-chiral patterns when we use the chiral palindrome loops. This is caused by the fact
that one uses only one type of fundamental control loop and its fundamental time reversed loop moving the particles perpendicular to the edge.
For an even chiral palindrome cyclotron loop, the motion upon entering the small lattice is on an adiabatic path.
The exit path is parallel to the entry path. Both paths are separated by one unit vector of the large lattice
and are thus equivalent. It follows that the exit is also adiabatic. The corresponding odd loops cannot adiabatically
connect the second minimum to the large lattice with the same loop segment. Hence, the exit to the large lattice
is of the ratchet type. Both exits, that of the forward cyclotron orbit and the time reverse exit, are again separated
by a unit vector of the large lattice which makes them equivalent. Therefore both ratchet jumps are in the same direction
and result in net edge transport.
 
\begin{figure}
	\includegraphics[width=0.95\columnwidth]{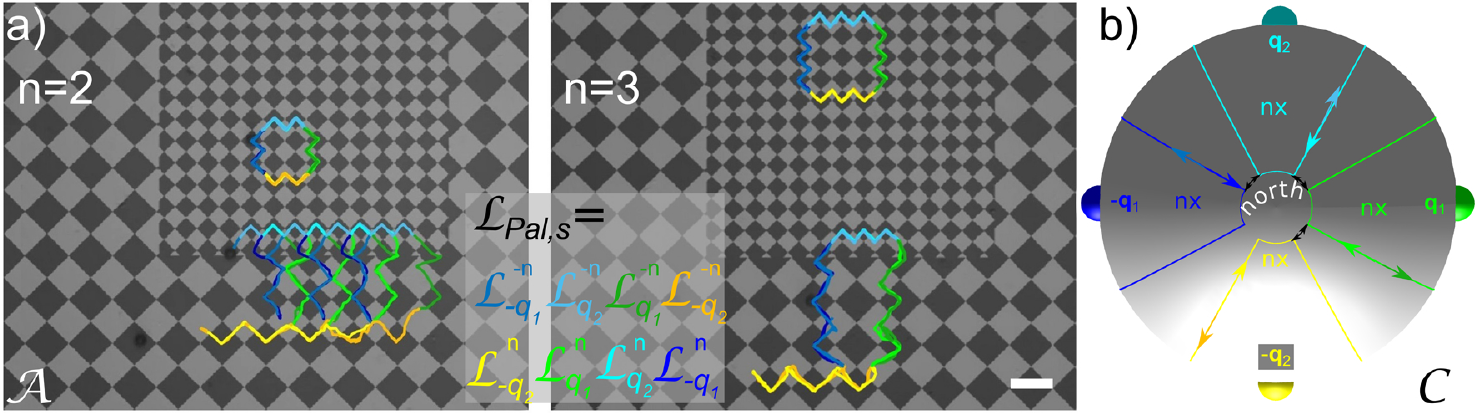}
	\caption{ {\bf a)} Experimentally determined bulk symmetric palindrome closed orbit and edge path of two colloidal particles on the square pattern in action space $\cal A$ for {\bf b)} the symmetric palindrome orbit ${\cal L}_{Pal,s}(n)$ in $\cal C$. The fundamental loops are shown in colors corresponding to the segments in ${\cal A}$. Fence points (encircled spheres) are shown in equivalent colors. The edge paths are closed for $n$ odd and open penetrating spirals for $n$ even. Fundamental loops start, end and are concatenated in the north in contrast to the examples shown in figs. 3-6. Scale bar is 14 $\mu m$.   Video clips of the motion of the paramagnetic colloidal particles are provided in $adfigure7\_2.avi$ -  $adfigure7\_3.avi$.}
	\label{evenmotion}
\end{figure}

If we drive the motion with a symmetric palindrome loop and the point of reversal lies within the large lattice
then the exit path of the small lattice is a mirror symmetric path to the entry path. Because the edge lacks
mirror symmetry the exit can be adiabatic or of the ratchet type for any $n$ odd or even. If it is adiabatic
then when the control loop reverses with the particle in the large lattice the reversed path must be adiabatic
as well at both entry and exit and the entire loop will cause a closed trajectory with no transport.
If on the other hand the exit is of the ratchet type we can use the same argument starting with the time reversed loop
to find that then also the second exit must be of the ratchet type. The direction of the ratchet jumps however
must not be, but can be, mirror images of each other because the edge is chiral. If we have a patch of the smaller lattice within the larger lattice there is an edge at the bottom of the patch and an opposite edge on the top of the patch. If we have edge transport
on one chiral edge using a symmetric palindrome loop then there cannot be edge transport on the opposite edge
because the reverse points of the same palindrome loop lie in large lattice for the first edge and in the small lattice for the opposite edge.    
In figure \ref{evenmotion} and in the video clips $adfigure7\_2.avi$ - $adfigure7\_3.avi$ we show the motion of
colloidal particles subject to a symmetric palindrome cyclotron modulation loop, where the transport is of the
ratchet type with net transport for $n=2$ (i.e. even) and of the adiabatic closed non-transport type for $n=3$ (i.e., odd).
Note that for this particular example the  fundamental modulation loops start, end, and are concatenated in the north of control space. 

The richness of edge transport phenomena and its connection to the symmetry of the pattern and the palindrome loops is shown in  in Fig. \ref{definitions},
where we summarize the resulting transport of different palindrome loops on chiral and symmetric patterns.

\rem{
\begin{table}
	\begin{tabular}{ | c | r | c | c | c |}
		\cline{1-5}
	\multicolumn{3}{|p{11cm}|} {\vbox{
			\center{\vspace{1mm{\bf palindrome loop with reversal point on large lattice}}}}}&\bf pattern&\bf edge transport/ character\\
			\cline{1-5}
&  &start of loop on non-alternating mirror symmetry plane&symmetric &no / adiabatic\\ \cline{3-5}
&symmetric  &start of loop on alternating mirror symmetry plane&symmetric  &no /  ratchet\\ \cline{3-5}
n even&  &&chiral &sometimes /adiabatic or ratchet\\ \cline{2-5}
&chiral  &&any&no / adiabatic \\ \cline{1-5}
&  &start of loop on non-alternating mirror symmetry plane&symmetric &no / ratchet\\ \cline{3-5}
n odd& symmetric &start of loop on alternating mirror symmetry plane&symmetric  &no /  adiabatic\\ \cline{3-5}
&  &&chiral &sometimes / adiabatic or ratchet\\ \cline{2-5}
&chiral  &&any&yes / ratchet\\ \cline{1-5}\hline\cline{1-5}
	\multicolumn{3}{|p{11cm}|} {\vbox{
		\center{\vspace{1mm{\bf palindrome loop with reversal point on small lattice}}}}}&\bf pattern&\bf edge transport\\ \cline{1-5}
	\multicolumn{3}{|p{8cm}|}{\centering{any}}&any&no/adiabatic	\\
\cline{1-5}
			\multicolumn{5}{c} {\bf table 1}\\
	\end{tabular}
\caption{\red{This table (if we need it) requires a caption. If we need the table we could add sketch of the patters and the loops???}}
\end{table}
}

\section{non-generic edges}

The patterns analysed here consist of two square lattices joined at an edge. There exists three degrees of freedom when designing
such patterns. Two of them are the positions at which we truncate each lattice in the direction perpendicular to the edge. The third
one is a relative shift between both lattices in the direction parallel to the edge. The full parameter space is therefore three dimensional.
The edge motion we have described in the preceding sections holds for most of this parameter space. In some regions of parameter space we can 
also suppress any adiabatic transport of particles from the larger lattice to the smaller. 
In such cases, Brownian dynamics simulation show that the generic situation is that ratchet jumps do not occur in the direction of the smaller lattice but do occur from the smaller toward the larger lattice. 
These ratchets are probably homotopic to some bulk adiabatic motion~\cite{tp1} of a nearby pattern. 

\section{Discussion}

In previous work \cite{colloidalTI}, we have shown that edge transport between topologically distinct lattices occurs in the form
of skipped orbits~\cite{Beenakker,Davies,Montambaux,Shi,Zhirov,Mancini}. Like the penetrating spiral orbits shown here, both forms
of edge transport are truly topological and the transport is determined and protected by topological invariants. Here,
the topological invariants are the winding numbers of the modulation loops around the fence points in control space.
Skipped orbits however differ from the penetrating spirals in two ways: First, skipped orbits propagate on one side of the edge only
and do not penetrate the other topologically distinct lattice. Second, skipped orbits fail to perform a displacement
command by one unit vector of the modulation loop, that is performed by colloidal particles in the bulk. That is, they really skip one command.
Penetrating spirals perform each command of the modulation loop either in the proper way of the lattice currently occupied
or in a particular edge penetrating move that may or may not differ from the corresponding bulk motion.
Skipped orbits of colloids are the analogue to topologically protected edge waves. Those edge waves due
to the topological contrast between two lattices obey a bulk-boundary correspondence. One can predict the number of skipped orbits
from the bulk orbits using the difference in the number of the fence points of both lattices encircled by a particular control loop in control space.
Such prediction however seems to be absent for the penetrating spirals propagating at the edge between topologically equivalent lattices since
both lattices share the same bulk control space.

It is clear that none of the edge modes discussed here occur along edges of matching square lattices with equal lattice constants.
In such cases the edges cannot be distinguished from the bulk because if we perfectly match the connection of both lattices the pattern
cannot be distinguished from a single lattice. 
The non-existence of edge states is the translation of the non-existence of chiral and propagating edge waves between topologically
equivalent wave supporting lattices. The difference of Chern numbers and of winding numbers of evolution operators
between topologically equivalent wave systems vanishes and so does the number of edge waves. We have performed also experiments on
edge transport between lattices with different ratios of unit cell sizes. When driving a system with cyclotron control loops for
ratios different from unity and for any direction of the edge we find propagating chiral edge modes.
It is therefore clear that a bulk-boundary correspondence cannot hold for particle systems at edges of differently sized lattices.
The failure of the bulk-boundary correspondence has been reported also in non-hermitian Hamiltonian systems
due to the presence of exceptional points for which the energy levels are complex and degenerated~\cite{Xiong}.
Kunst et al.~\cite{Kunst} reported on a generalization of the bulk-boundary correspondence that includes non-hermitian Hamiltonians.
Our example however proves the impossibility of a bulk-boundary correspondence principle for particle systems between topologically
equivalent lattices of different scale. 
It might also be possible that a break down of the bulk-boundary correspondence occurs for wave systems at edges between differently sized unit cells. 



\section*{Acknowledgements}
A. T. is supported by PhD fellowship of the University of Kassel. Publication is supported by the University of Bayreuth open access fund. 


\section{Appendix}
We use Brownian dynamics simulations to study the motion of isolated point dipoles moving in the total colloidal potential given by
 	\begin{equation}\label{V}
V({\bf x}_{\cal A},{\bf H}_{\textrm{ext}}(t))\propto -{\bf H}_{\textrm{ext}}(t)\cdot {\bf H}_p({\bf x}_{\cal A}),
 \end{equation}
 where ${\bf H}_{\textrm{ext}}(t)$ is the external field at time $t$, and ${\bf H}_p({\bf x}_{\cal A})$ is the magnetic
 field created by the pattern at position ${\bf x}_{\cal A}$ in $\cal A$. The pattern field is calculated as 
 	\begin{equation}
{\bf H}_p({\bf x}_{\cal A})=\int d^2 {\bf x}_{\cal A}' \frac{({\bf x}_{\cal A}-{\bf x}_{\cal A}')+z{\bf e}_z }{4\pi|({\bf x}_{\cal A}-{\bf x}_{\cal A}')+z{\bf e}_z|^3} M_z({\bf x}_{\cal A}')
\end{equation}
 where $M_z({\bf x}_{\cal A}')$ is the magnetization pattern and $z\approx 0.2 a$ is the elevation of the particles above the pattern with large lattice constant a.
 The equation of motion in the overdamped limit is
	\begin{equation}
	{\bf \xi\dot x_{\cal A}}(t)=-\nabla_{\cal A} V({\bf x}_{\cal A},{\bf H}_{\textrm{ext}}(t))+\eta(t),
	\end{equation}
	where $\xi$ is the friction coefficient, and $\eta$ is a Gaussian random force with a variance given by
	the fluctuation-dissipation theorem. The equation of motion is integrated in time with a standard Euler
        algorithm. We use a time step $T/dt\approx2\cdot10^5$ with $T$ the period of a modulation loop ${\cal L_C}$.

	From equation\ref{V} it follows that a point ${\bf x}_{\cal A}$ is made stationary by an external field 
	\begin{equation}
	{\bf H}_{\textrm{ext}}^s({\bf x}_{\cal A})\propto\pm\partial_1{\bf H}_p({\bf x}_{\cal A})\times\partial_2{\bf H}_p({\bf x}_{\cal A})
	\end{equation}
	The allowed (forbidden) regions are the regions in $\cal A$ of positive (negative) determinant
	  	\begin{equation}
       | {\bf H}_{\textrm{ext}}^s({\bf x}_{\cal A})\cdot \nabla_{\cal A}\nabla_{\cal A}{\bf H}_p({\bf x}_{\cal A})|
	 \end{equation}
	 where the partial derivatives are taken along arbitrarily chosen coordinates in $\cal A$.

\section*{Competing interest }The authors declare no competing financial interests

 \emph{}

\end{document}